# The ground state in a proximity to a possible Kitaev spin liquid:

## An undistorted honeycomb iridate Na$_x$IrO$_3$ (0.60 ≤ x ≤ 0.80)

Hengdi Zhao[1], Bing Hu[1,2], Feng Ye[3], Minhyea Lee[1], Pedro Schlottmann[4] and Gang Cao[1*]

[1]Department of Physics, University of Colorado at Boulder, Boulder, Colorado 80309, USA

[2]School of Mathematics and Physics, North China Electric Power University, Beijing 102206, China

[3] Neutron Scattering Division, Oak Ridge National Laboratory, Oak Ridge, TN 37831, USA

[4] Department of Physics, Florida State University, Tallahassee, FL 32306, USA

We report results of our study of a newly synthesized honeycomb iridate Na$_x$IrO$_3$ (0.60 ≤ x ≤ 0.80). Single-crystal Na$_x$IrO$_3$ adopts a honeycomb lattice noticeably without distortions and stacking disorder inherently existent in its sister compound Na$_2$IrO$_3$. The oxidation state of the Ir ion is a mixed valence state resulting from a majority Ir$^{5+}$(5$d^4$) ion and a minority Ir$^{6+}$(5$d^3$) ion. Na$_x$IrO$_3$ is a Mott insulator likely with a predominant pseudospin = 1 state. It exhibits an effective moment of 1.1 $\mu_B$/Ir and a Curie-Weiss temperature of -19 K but with no discernable long-range order above 1 K. The physical behavior below 1 K features two prominent anomalies at $T_h$ = 0.9 K and $T_l$ = 0.12 K in both the heat capacity and AC magnetic susceptibility. Intermediate between $T_h$ and $T_l$ lies a pronounced temperature linearity of the heat capacity with a large slope of 77 mJ/mole K$^2$, a feature expected for highly correlated metals but not at all for insulators. These results along with comparison drawn with the honeycomb lattices Na$_2$IrO$_3$ and (Na$_{0.2}$Li$_{0.8}$)$_2$IrO$_3$ point to an exotic ground state in a proximity to a possible Kitaev spin liquid.

*gang.cao@colorado.edu

***Introduction*** Honeycomb lattices with strong spin-orbit interactions (SOI) have been extensively sought and studied because they are most desirable as possible realization of the exactly solvable spin liquid model developed by Kitaev [1]. The honeycomb lattices feature $MO_6$ (M = Ir or Ru) octahedra that are edge-sharing with 90° M-O-M bonds. The magnetic exchange is anisotropically bond-dependent. When individual spins at the sites of a honeycomb lattice are restricted to align along any one of the three bond directions, the Kitaev model predicts a quantum spin-liquid ground state. This novel state hosts short-range correlations and the spin degrees of freedom that fractionalize into Majorana fermions in static $Z_2$ gauge fields. Theoretical treatments of the honeycomb lattices $Na_2IrO_3$ and $Li_2IrO_3$ (including β and γ phases) and more recently α-$RuCl_3$ [e.g., 2-8] have inspired a large body of experimental work that anticipates the Kitaev physics [e.g., 9-30]. Although strongly frustrated, all these honeycomb lattices are antiferromagnetically ordered with the Néel temperature ranging from 7 K to 18 K. There has been no clear-cut materials realization of a quantum spin liquid (QSL) at ambient conditions thus far.

The absence of a QSL in these honeycomb lattices indicates that the Heisenberg interaction, which competes with the strong Kitaev interaction, is still consequential in the Kitaev-Heisenberg model [25], in part because of stacking disorder and distortions often characterized by unequal M-M bonds inherently existent in these honeycomb lattices [12, 15, 25-30]. This is an experimental challenge that is particularly daunting for honeycomb lattices with strong SOI that renders an extraordinary susceptibility of the ground state to the lattice degrees of freedom [31, 32].

The overwhelming balance of interest has been devoted to the honeycomb lattices hosting five *d*-electrons and a pseudospin = 1/2 state, such as $Na_2IrO_3$, $Li_2IrO_3$ and α-$RuCl_3$, as quantum fluctuations in a QSL with a pseudospin = 1/2 state are more resilient to classical effects. Honeycomb lattices with a higher spin state, such as the ruthenates $Na_2RuO_3$ and $Li_2RuO_3$ with a $Ru^{4+}(4d^4)$ ion and a spin = 1 state [33], have remained largely unexplored. It is encouraging that a recent study extends the search of QSLs to honeycomb materials with S = 3/2 that show a two-peak characteristic in the heat capacity, a promising sign of a QSL [34].

Here we report structural and physical properties of our newly synthesized single crystals of $Na_xIrO_3$ (0.60 ≤ x ≤ 0.80). One outstanding feature of this new compound is that $Na_xIrO_3$ adopts an undistorted honeycomb lattice without stacking disorder or intermixing of Na and Ir inherently existent in its sister compound $Na_2IrO_3$ [12]. Our chemical and structural analysis indicates that the oxidation state of the Ir ion is a mixed valence state resulting from a majority $Ir^{5+}(5d^4)$ ion and a minority $Ir^{6+}(5d^3)$ ion. The insulating $Na_xIrO_3$ shows an effective moment of 1.1 $\mu_B$/Ir, too large for an anticipated singlet ground state for a strong SOI limit, suggesting a predominant pseudospin = 1 state. It exhibits a Curie-Weiss temperature of -19 K with no discernable long-range order above 1 K. The physical behavior below 1 K presents two prominent anomalies at $T_h$ = 0.9 K and $T_l$ = 0.12 K, respectively, in both the heat capacity and AC magnetic susceptibility. Intermediate between $T_h$ and $T_l$ lies a pronounced temperature linearity of the heat capacity with an unexpectedly large slope of 77 mJ/mole $K^2$, a feature expected for highly correlated metals but not at all for any insulators. These results along with comparison drawn with $Na_2IrO_3$ and $(Na_{0.2}Li_{0.8})_2IrO_3$ point out an exotic ground state that hosts strong quantum fluctuations likely coexisting with a short-range spin order. Note that the ground state of $Na_xIrO_3$ is insensitive to x.

*Crystal structure* The crystal structure of the single crystal $Na_xIrO_3$ was determined independently using a Bruker D8 Quest ECO single-crystal diffractometer at the University of Colorado Boulder and a Rigaku XtaLAB PRO diffractometer at the Oak Ridge National Laboratory after thorough examinations of dozens of single crystal $Na_xIrO_3$. All datasets were refined by using APEX3 and/or SHELXL-2014 program [35] (see Ref. 36 for more details). Single crystal $Na_xIrO_3$ with x ranging between 0.60 and 0.80 adopts a quasi-two-dimensional hexagonal structure with space group *P-31m* (No. 162) (**Fig.1**). There are two distinct Na sites, Na1 and Na2 (**Figs.1a-b**). The Na1 site resides at the center of the honeycomb ring in the *ab* or honeycomb plane whereas the Na2 site exists between the honeycomb planes. Almost all Na vacancies occur at the Na1 sites within the honeycomb rings whereas the Na2 sites are fully or nearly fully occupied (**Figs.1a-b**); this explains that x in $Na_xIrO_3$ is hardly smaller than 0.50. Most importantly, the edge-sharing $IrO_6$ octahedra form a robust, *undistorted* honeycomb lattice characterized by an exactly equal Ir-Ir bond distance, $d_{Ir-Ir}$, between all neighboring Ir atoms (**Fig.1c**), independent of the Na1 deficiency [36]. The Ir sites are fully or nearly fully occupied, and there is no discernable oxygen deficiency. The characteristic of the honeycomb lattice is also evident in both the X-ray diffraction pattern (**Fig.1d**) and the habit of the $Na_xIrO_3$ crystals (**Fig.1e**).

It is remarkable that intermixing of the Na1 and Ir sites in the stoichiometric $Na_2IrO_3$ is a common occurrence and accounts for stacking disorder and the distorted honeycomb lattice indicated by two distinct Ir-Ir bond distances, a long one (3.073 Å) and a short one (3.071 Å) [12], both of which are shorter than that in $Na_xIrO_3$. In contrast, there is no intermixing of the Na1 and Ir sites in $Na_xIrO_3$, giving rise to a perfect honeycomb lattice. However, like those in $Na_2IrO_3$ [12], the $IrO_6$ octahedra in $Na_xIrO_3$ undergo a compression along the *c* axis. The O-Ir-O bond angle related to the shared edge of the neighboring octahedra is reduced to 80.2° from the undistorted 90° (compared to 84.1° and 84.5° in $Na_2IrO_3$) and the rest of the O-Ir-O bond angles is increased to 93.4° accordingly (**Fig.1c**). Such a trigonal crystal field could have implications for the splitting of $J_{eff}$ = 3/2 bands [37].

In terms of the oxidation state of Ir in $Na_xIrO_3$, for x = 1, the Ir ion will be pentavalent $Ir^{5+}$. However, the average value of x among dozens of the examined crystal samples is around 0.70, and this gives rise to an average oxidation state of $Ir^{5.3+}$, a mixed valence state resulting from 70% $Ir^{5+}(5d^4)$ and 30% $Ir^{6+}(5d^3)$. The determination of an $Ir^{5+}$ majority in $Na_xIrO_3$ is also consistent with the following analysis. The Ir-O bond distance, $d_{Ir-O}$, is 2.028Å in $Na_xIrO_3$, expectedly shorter than 2.188 Å in $Na_2IrO_3$ with $Ir^{4+}(5d^5)$; the corresponding ratio of $d_{Ir-O}(Na_2IrO_3)$ to $d_{Ir-O}(Na_xIrO_3)$, $R_{Ir-O}$, is 1.079. The difference in $d_{Ir-O}$ is a result of the difference in the ionic radius, $r_{Ir}$, of Ir, which decreases significantly from 0.625 Å to 0.570 Å to 0.521 Å for $Ir^{4+}$, $Ir^{5+}$ and $Ir^{6+}$, respectively, as more electrons are removed from the Ir ion. The ratio of $r_{Ir4+}$ to $r_{Ir5+}$ or $R(r_{Ir4+}/r_{Ir5+}) = 1.097$ and $r_{Ir4+}$ to $r_{Ir6+}$ or $R(r_{Ir4+}/r_{Ir6+}) = 1.200$. It is apparent that $R_{Ir-O}$ (=1.079) is much closer to $R(r_{Ir4+}/r_{Ir5+})$ than to $R(r_{Ir4+}/r_{Ir6+})$, supporting that Ir is predominately pentavalent $Ir^{5+}$ in $Na_xIrO_3$. Unless specified, *all data presented here are those for x ≈ 0.70.*

It is emphasized that our close examination of single-crystal samples with x ranging from 0.60 to 0.80 indicates that structural and physical properties of $Na_xIrO_3$ are insensitive to x or Na deficiency [36]. $Na_xIrO_3$ thus sharply contrasts with another Na-deficient compound, $Na_xCoO_2$, in which x varies widely from 0.3 to 0.80 and whose ground state drastically evolves with x [38, 39].

***Physical properties*** $Na_xIrO_3$ is a Mott insulator. The *ab*-plane electrical resistivity, $\rho_{ab}$, rises by five orders of magnitude in a manner consistent with a variable-range hopping of carriers between localized states as temperature, T, decreases from 380 K to 25 K (**Fig.2a**).

The low-field magnetization for both the *ab* plane and *c* axis, $M_{ab}$ and $M_c$, exhibits no anomaly in the interval 1.8 K - 350 K at $\mu_oH = 0.5$ T (**Figs.2b-c**). A Curie-Weiss analysis of the magnetic susceptibility, $\chi$, yields an effective moment, $\mu_{eff}$, of 0.9 and 1.1 $\mu_B$/Ir and a Curie-Weiss temperature, $\theta_{CW}$, of -19 and -15 K for the *ab* plane and the *c* axis, respectively (**Fig.2c**; $\Delta\chi = \chi - \chi_o$, with $\chi_o$ the T-independent susceptibility). The values of $\mu_{eff}$ are essentially identical to those of the double perovskite antiferromagnets $Sr_2YIrO_6$ and $Ba_2YIrO_6$ with pentavalent $Ir^{5+}(5d^4)$ ions [40, 41]. These values are clearly too large for a singlet $J_{eff} = 0$ state anticipated for a strong SOI limit in iridates with $Ir^{5+}(5d^4)$ ions but considerably smaller than 2.83 $\mu_B$/Ir expected for a spin-only S = 1 state without SOI. A reduced value of $\mu_{eff}$ is commonplace in iridates, in part because the strong SOI causes a partial cancellation of the spin and orbital contributions [32]. Nevertheless, despite the sizable $\mu_{eff}$ and $\theta_{CW}$, no long-range magnetic order occurs above 1.8 K, indicating overwhelming quantum fluctuations in the honeycomb lattice and calling for an examination of the ground state below 1.8 K.

The AC magnetic susceptibility, $\chi_{ac}$, and the heat capacity, C(T), are thus measured down to 0.05 K. $\chi_{ac}$ at DC field H = 0 displays two peaks denoted by $T_h$ and $T_l$ (red curve in **Fig. 3a**), namely, $T_h$ = 0.9 K and $T_l$ = 0.12 K for a broad peak and a sharper peak in $\chi_{ac}$, respectively. $T_h$ and $T_l$ track two prominent anomalies observed in C(T) (blue curve, right scale in **Fig.3a**). C(T) exhibits a broad and yet visible peak near $T_h$ and an abrupt rise at $T_l$. The entropy removal, $\Delta S$, also shows a slope change near both $T_h$ and $T_l$, respectively, and a noticeable shoulder situated between $T_h$ and $T_l$ (**Fig.3b**). $\Delta S$ is estimated to be 0.11 J/mole K. This value of $\Delta S$, which is comparable to that for the quantum liquid $Ba_4Ir_3O_{10}$ [42] but much smaller than the Rln(3) = 9.12 J/mole K expected for an S = 1 state, implies that $Na_xIrO_3$ behaves like a Fermi liquid metal where most of the entropy removal happens near a Fermi temperature, $T_F$, and the T-linear C(T) occurs at T << $T_F$.

Indeed, C(T) exhibits a pronounced linear temperature dependence between $T_h$ and $T_l$ or C(T) = $\gamma$T with an unusually large coefficient $\gamma$ = 77 mJ/mole $K^2$ (**Fig.3c**). This behavior is anticipated for highly correlated metals and not at all expected for conventional insulators. The linear heat capacity suggests gapless excitations, and the value of $\gamma$ implies a large residual entropy despite such low temperatures.

For comparison and contrast, C(T) of $Na_2IrO_3$ is also measured and illustrated along with that of $Na_xIrO_3$ in **Fig.3c**. The starkly different C(T) of the two sister compounds may help rule out a possible nuclear Schottky anomaly, supporting the unique nature of the upturn marked by $T_l$. This point is further strengthened by the corresponding anomaly near $T_l$ in $\chi_{ac}$. Indeed, a nonlinear behavior of C(T) in a plot of C(T) vs $T^{-2}$ is inconsistent with that of the nuclear Schottky anomaly [Fig.3 in Ref. 36] because the heat capacity of a nuclear Schottky anomaly is expected to scale with $T^{-2}$. It is also remarkable that C(T) of $Na_2IrO_3$ (brown curve in **Fig.3c**) monotonically approaches zero with decreasing T due to magnetic entropy removal at $T \leq T_N$. In contrast, C(T) of $Na_xIrO_3$ (blue curve in **Fig.3c**) is much larger in general and reaches 25 mJ/mole K at $T_l$ before rising to 68 mJ/mole K at 0.05 K, highlighting strong quantum fluctuations even at sub-Kelvin temperatures.

The characteristic of C(T) for $Na_xIrO_3$ seems resilient against magnetic fields comparable to the energy scale of $T_h$ and $T_l$. As shown in **Figs.4a-b**, the temperature dependence of C(T) changes only slightly at $\mu_oH = 1$ T but more significantly at $\mu_oH = 3$ T. The C/T vs T plot in **Fig.4b** illustrates an increasing separation between $T_h$ and $T_l$ with increasing H. However, application of stronger magnetic field, such as 14 T, suppresses both $T_h$ and $T_l$ and removes residual entropy, resulting in a behavior consistent with that of a conventional insulator (**Fig.4a**).

Such unusual thermal behavior at sub-Kelvin temperatures clearly contrasts with that of $Na_2IrO_3$ with $T_N$ = 18 K (**Fig.3c**) but bears certain resemblance to that of $(Na_{1-x}Li_x)_2IrO_3$ with x = 0.80, at which $T_N$ = 1.4 K [15] (**Fig.5a**). This comparison is revealing. An early study of $(Na_{1-x}Li_x)_2IrO_3$ demonstrates that $T_N$ is initially suppressed from 18 K for x = 0 to 5 K for x = 0.28 and then to 1.2 K for x = 0.70 and 1.4 K for x = 0.80 before it rises to 7 K for x = 0.90 [15] (**Inset** in **Fig.5b**). Furthermore, the honeycomb structure near x=0.70 and 0.80 is least distorted [15], leading to a speculation that $(Na_{1-x}Li_x)_2IrO_3$ with x=0.70 and 0.80 may be closest to the spin liquid [15, 43, 44]. As shown in **Fig.5a**, C(T) at T > 0.9 K for both $Na_xIrO_3$ and $(Na_{0.2}Li_{0.8})_2IrO_3$ behaves in a similar manner, suggesting a similar magnetic nature. However, C(T) for $(Na_{0.2}Li_{0.8})_2IrO_3$ undergoes a rapid decrease below $T_N$ = 1.4 K, approaching zero at 0.05 K owing to the magnetic entropy removal; in contrast, C(T) for $Na_xIrO_3$ decreases less rapidly below $T_h$ (= 0.9 K) reaching a minimum of 25 mJ/mole K at $T_l$ before it abruptly rises below $T_l$ (**Fig.5b**). The comparison suggests that $T_h$ may be associated with a short-range order rather than a long-range order because a large residual entropy below $T_h$ remains; $T_l$ marks an onset of strong quantum fluctuations which increases as T approaches absolute zero.

The peculiar heat capacity of $Na_xIrO_3$ invokes certain theoretical arguments. Theoretical studies of thermal properties for the Kitaev model predict two peaks in the temperature dependence of the specific heat for honeycomb lattices [34, 43, 44]. This two-peak characteristic is a result of fractionalizing a single quantum spin into two types of Majorana fermions, namely, the itinerant Majorana fermion and the localized Majorana fermion. The two peaks in the heat capacity thus correspond to the onset of the thermal excitations or short-range spin correlations of the itinerant Majorana fermions at the high-temperature peak and the thermal excitation of the localized Majorana fermions at the low-temperature peak, respectively [35, 43, 44]. The theoretical studies also anticipate a linear temperature dependence of the heat capacity between the two peaks [43] and a half-plateau-like temperature dependences of the entropy between the two peaks due to the thermal fractionalization of the spin degrees of freedom [44].

It is particularly intriguing that an array of the observed phenomena - the two anomalies marked by $T_h$ and $T_l$ in C(T) (**Fig.3a**), the linearity of C(T) along with the large γ (**Fig.3c**) and the shoulder of ΔS between $T_h$ and $T_l$ (**Fig.3b**) - suggests a strong relevance of $Na_xIrO_3$ to the theoretical anticipation for a QSL and an exotic ground state that hosts strong quantum fluctuations coexisting with a short-range spin order. These results along with the comparison with $Na_2IrO_3$ and $(Na_{0.2}Li_{0.8})_2IrO_3$ inspire a speculation that such a ground state may be in a proximity to the Kitaev spin liquid. Certainly, with a perfect honeycomb lattice and a predominant pseudospin = 1 state $Na_xIrO_3$ provides a new, perhaps unique candidate material for the search of a Kitaev QSL, which has been elusive to date.

**Acknowledgement** This work is supported by NSF via grant DMR 1903888.

**Captions**

**Fig.1. Crystal Structure of Na$_x$IrO$_3$: (a)** the *ab* plane, **(b)** *ab* planes stacking along the *c* axis, **(c)** the honeycomb ring formed by edge-sharing IrO$_6$ octahedra (the values are for x=0.73), **(d)** a snapshot of the X-ray diffraction pattern showing the honeycomb lattice and **(e)** a single-crystal sample with a hexagon. Note that VESTA [45], a virtualization program for crystal structures used here, yields mixed colors of both yellow and white for Na1 sites within the honeycomb rings to indicate the Na deficiency and a single color of yellow for Na2 sites to illustrate the full occupancy. More importantly, all Ir-Ir bond distances in the honeycomb ring are equal, giving rise to a robust, undistorted honeycomb lattice, regardless Na vacancies at the Na1 sites.

**Fig.2. Transport and Magnetic Properties of Na$_x$IrO$_3$:** The temperature dependence of **(a)** the *ab*-plane resistivity $\rho_{ab}$, (b) the magnetization for the *ab* plane and *c* axis, M$_{ab}$ and M$_c$, and (c) the reciprocal magnetic susceptibility for the *ab* plane and *c* axis $\Delta\chi_{ab}^{-1}$ and $\Delta\chi_c^{-1}$. **Inset** in **(b)**: M$_{ab}$ and M$_c$ vs log T for clarification.

**Fig.3. AC Susceptibility and thermal properties of Na$_x$IrO$_3$:** The temperature dependence of **(a)** the AC susceptibility $\chi_{ac}$ (red curve) at 10 kHz and an AC field H$_{ac}$ = 3.1 Oe, and heat capacity C(T) (blue curve, right scale), **(b)** the entropy removal $\Delta$S and **(c)** C(T) for both Na$_x$IrO$_3$ and Na$_2$IrO$_3$ for comparison. Note that the yellow oval in (a) outlines the board peak near T$_h$ and the red dashed lines in (b) and (c) are a guide to the eye, highlighting the linearity of the region between T$_h$ and T$_l$.

**Fig.4. Heat Capacity of Na$_x$IrO$_3$:** The temperature dependence of **(a)** C(T) and **(b)** C/T at a few representative magnetic fields applied along the *c* axis.

**Fig.5. Heat Capacity of Na$_x$IrO$_3$ and (Na$_{0.2}$Li$_{0.8}$)$_2$IrO$_3$ for comparison:** The temperature dependence of **(a)** C(T) and **(b)** C/T. **Inset** in **(a)**: the *ab*-plane magnetic susceptibility $\chi_a$ for (Na$_{0.2}$Li$_{0.8}$)$_2$IrO$_3$; **Inset in (b):** T$_N$ as a function of Li concentration x. Note that T$_N$ = 1.4 K at x=0.80 of Li doping where the honeycomb lattice is nearly undistorted [15].

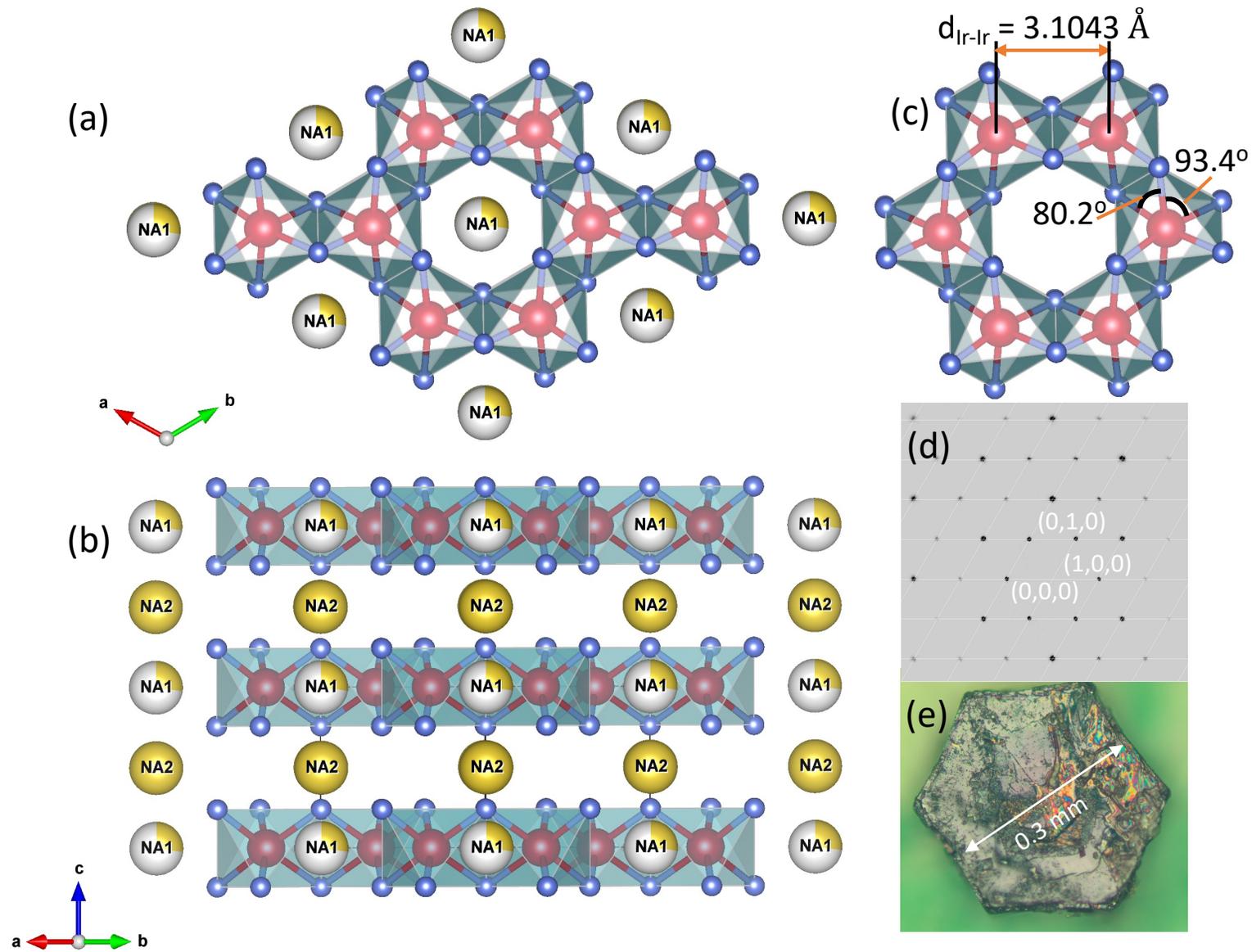

Figure 1

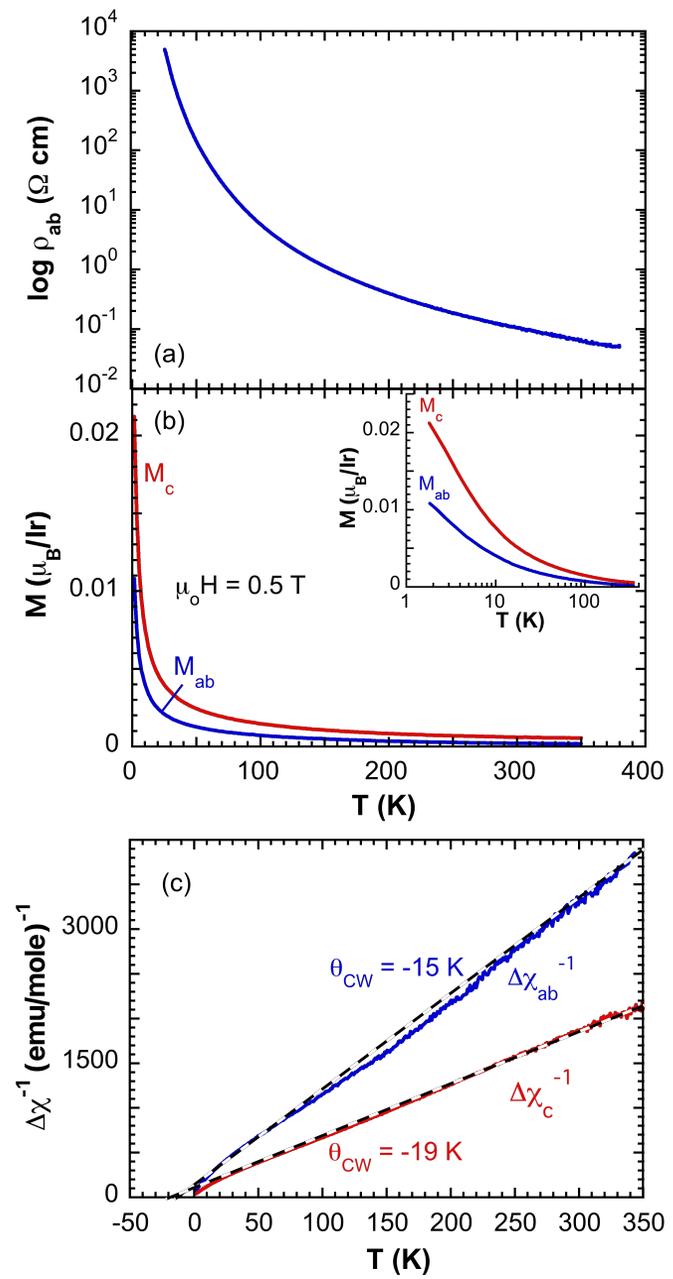

Figure 2

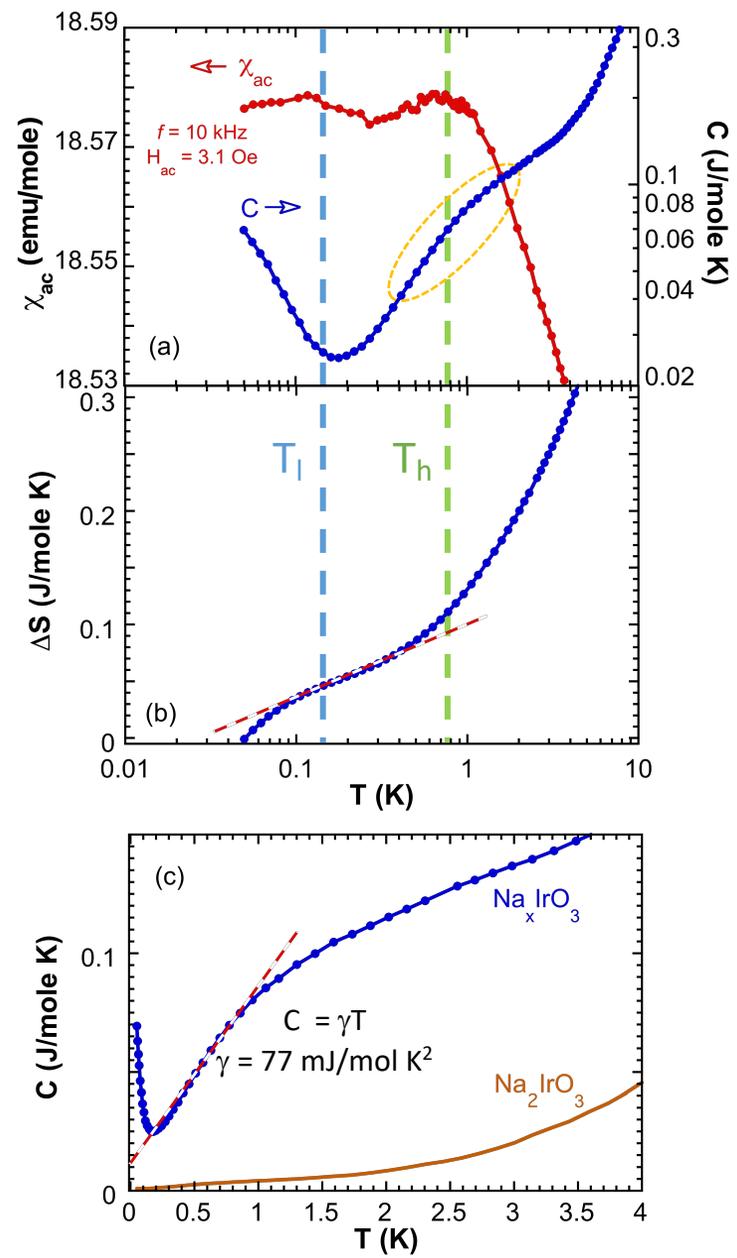

Figure 3

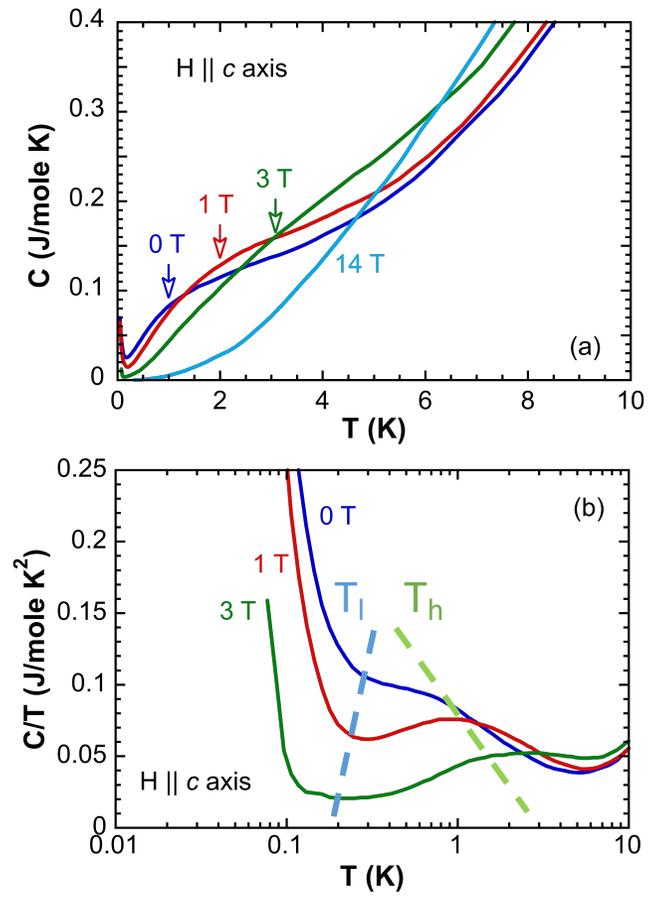

Figure 4

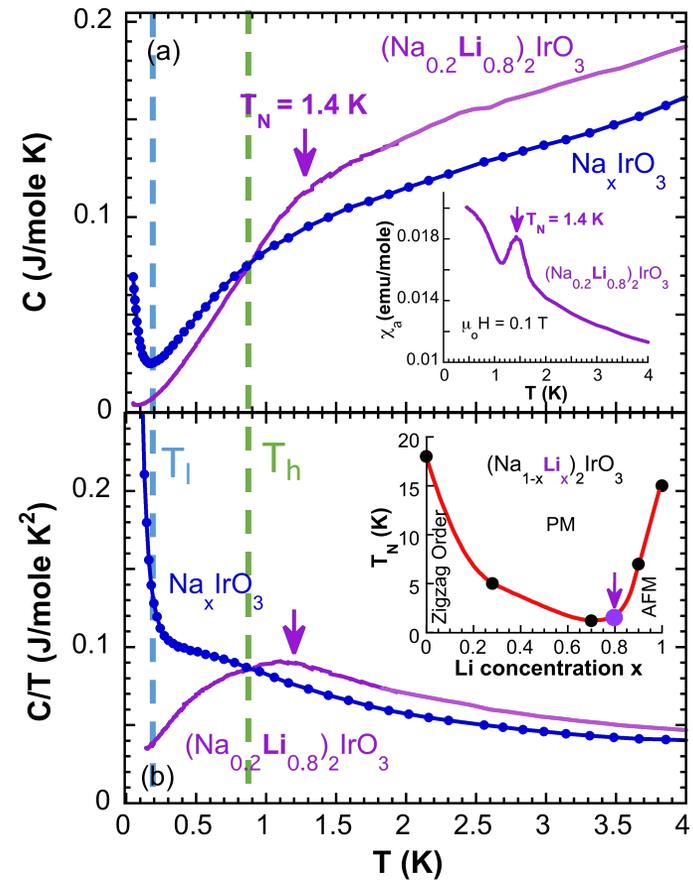

Figure 5